# A study of the magnetotransport properties of the graphene (I. Monolayer)

M. A. Hidalgo

Departamento de Física y Matemáticas

Universidad de Alcalá

Alcalá de Henares, Madrid, Spain

Correspondence and request for materials should be addressed to miguel.hidalgo@uah.es

**Abstract**

We present a single electron approach to analyse the magnetotransport properties of the monolayer graphene as a function of both, the gate voltage and the magnetic field; and, also, their evolution with temperature. The model proposed means the extension of our previous one developed for studying the quantum Hall and Shubnikov-de Haas effects of a two-dimensional electron system in a semiconductor quantum well. Now, the study in this framework of both phenomena in graphene involves including the presence of two bands and two degeneracy valleys, (points K and K' in the reciprocal space). Based in a single electron approximation, we show it is capable to reproduce the entire characteristics observed in the experiments for the Hall and diagonal magnetoconductivities (and the corresponding magnetoresistivities), as a function of the gate voltage and the magnetic field. In the model the observed Hall plateaux series in the monolayer graphene, determined by the expression $\pm 2(2n+1)$, arises in a natural way as a consequence of the particular quantization of the energy spectrum of graphene. Therefore, on the other hand the proposed approach integrates the quantum Hall effects observed in graphene and quantum wells semiconductors.

1. **Introduction**

The discover in a two-dimensional electron system (2DES) embedded in a semiconductor quantum well (QW) of the integer quantum Hall effect (IQHE) in 1980 [1,2], and the fractional quantum Hall effect (FQHE) two years later, in 1982 [3], were a milestone not only in solid state physics for their intrinsic interest from a fundamental point of view, but for their potential technological implications. Although the last years the study of these phenomena becomes stagnant, however, the observation of likeness effects in graphene has reactivated the interest on the magnetotransport properties of 2DES. The characteristics of them in both kind of electron systems are similar: minima or zeroes in the longitudinal resistance, i.e. Shubnikov-de Haas effect (SdH), and well-defined plateaux in the Hall resistance (at integer or fraction values of the fundamental Hall resistance, $R_H = h\ e^2$ ).

On the other hand, the IQHE, measured in any 2DES in a semiconductor QW, shows plateaux following the integer series $2n$, with values 2, 4, 6, 8…, when there is spin degeneration; or $n$, i.e., 1, 2, 3, 4, 5…, if that spin degeneration is broken [4]. On the other hand, the FQHE plateaux shows an extensive series of fractional values, mainly with odd denominators: {1/3} [2], {4/5, 2/3, 3/5, 4/7, 5/9, 5/3, 8/5, 11/7, 10/7, 7/5, 4/3, 9/7} [5]; {5/3, 8/5, 10/7, 7/5, 4/3, 9/7, 4/5, 3/4, 5/7, 2/3, 3/5, 4/7, 5/9, 6/11, 7/13, 6/13, 5/11, 4/9, 3/7, 2/5, and also 8/3, 19/7, 33/13, 32/13, 7/3, 16/7} [6,7]; {2/3, 7/11, 3/5, 4/7, 5/9, 6/11, 5/11, 4/9, 3/7, 2/5, 1/3, 2/7, 3/11, 4/15, 3/13, 2/9} [8]; {14/5, 19/7, 8/3, 13/5, 23/9, 22/9, 17/7, 12/5, 7/3, 16/7, 11/5} [9]; {19/5, 16/5, 14/5, 8/5, 7/3, 11/5, 11/3, 18/5, 17/5, 10/3, 13/5, 12/5} [10]; although can also be observed with even denominator: {15/4, 7/2, 13/4, 11/4, 5/2, 9/4} [6]; {11/4, 5/2, 9/4} [7]; {1/4} [8]; {11/4, 21/8, 5/2, 19/8, 9/4} [9]; {7/2, 5/2} [10].

While the IQHE is observed in 2DESs embedded in any QW, (MOSFET or semiconductor heterostructure), the observation of the FQHE requires high motilities heterostructures (for example, GaAs-AlGaAs). From the theoretical point of view, both phenomena, IQHE and

FQHE, have been tried to understand in different scenarios: in the first case as the effect of the localization of the electrons in the 2DES, considering it as an electron gas; and as a consequence of the Fermi liquid behaviour of the 2DES [11,12], in the second case. However, in both views several crucial points remain unexplained [15]. In particular, the recent observation of an integer quantum Hall effect in graphene leads to revise the theoretical explanation of the IQHE such as was established thirty years ago by Laughlin [2], and based on electron localization effects.

On the other hand, an alternative global approach for the IQHE and FQHE in semiconductors in the context of the single particle approximation has recently been presented [4,13,14,15].

So, the amazing discover of the quantum Hall effect in graphene [16,17], (GQHE), -in some references is called anomalous Hall quantum Hall effect-, arises the interest to revisit all the Hall effects.

The GQHE in the monolayer graphene (MLG), (a single two-dimensional sheet of carbon atoms in a honeycomb lattice) [17,18], is characterized for the observation of a series of quantum Hall plateaux at the integer values ±2, ±6, ±10, ±14, ±18, ±22…, i.e. following the succession $\pm 2(2n+1)$. Afterward the publication of these results, in bilayer graphene (BLG), (two sheets of carbon atoms), was also observed a different series of Hall plateaux, at the integer values 0, ±4, ±8, ±12, ±16, ±20, ±24…, i.e., sequence summarized in the expression $\pm 4n$ [19].

As expected, these experimental results have lead to a rapid increase in the number of publications devoted to such easily fabricated materials, for their intrinsic interest from the fundamental point of view and, overall, for their potential implications in technology [20,21,22]. By the way, there has been an extraordinary progress in the development of a new quantum standard for resistance [23,24]. The possibilities based on graphene are immense, and extensive backgrounds and excellent reviews on its physics can already be found in literature [20,21,22].

The structure of the paper is the following: in the next section we determine the density of states for the 2DES in MLG, basic to obtain the magnetotransport magnitudes, which we calculate in Section 3. In the fourth section we describe the results of the model, showing the possibilities it provides in the analysis of all the aspects of those phenomena. Finally, we have added a summary and discussion section, Section 5.

## 2. The density of states for the monolayer graphene

MLG is a flat layer of carbon atoms arranged in a hexagonal lattice with two carbon atoms per unit cell. Of the four valence states, three $sp^2$ orbitals form a $\sigma$ state with three neighbouring carbon atoms, and one $p$ orbital develops into delocalized $\pi$ and $\pi^*$ states that provide the occupied valence band (VB) and the lowest unoccupied conduction band (CB). The $\pi$ and $\pi^*$ states are degenerate at the corner of the hexagonal Brillouin zone. This degeneracy occurs at the Dirac crossing energy [21].

The undoped graphene is a zero band gap semimetal, with the Fermi level located at the intersection between the valence and the conduction bands, lying at $E=0$, and all the states with $E<0$ are filled while those with $E>0$ are empty. In the reciprocal space that intersection is found at the six corners of the first Brillouin zone. These six corners fall into two groups of three equivalent points, differing only by a reciprocal lattice vector. These two groups represent two non-equivalent Dirac points (DP), K and K'. [20,21,22]

We now consider $\varepsilon_{VB}$ and $\varepsilon_{CB}$ as the corresponding Hilbert spaces associated with the dynamic states of each band in each Dirac point, (and that are the same for each one, K and K').

Because we have two symmetric bands we can assume that the effective masses are the same, i.e., $m_{VB}^* = m_{CB}^* = m^*$, where $m_{VB}^*$ and $m_{CB}^*$ correspond to the effective masses of the valence and conduction bands, respectively. [22,25]

As we have mentioned above, our approach to the magnetotrasport properties of the graphene will be based on the same steps given in our previous work for the IQHE and FQHE [14,15], -i.e., within a single electron approximation-, although now considering the energy spectrum of electrons in graphene under the application of a magnetic field.

Thus, we have first to determine the quantized electron states and for obtaining them we take into account the symmetric gauge $\mathbf{A} = B(-y,x)/2$, being $\mathbf{B} = (0,0,B)$ the magnetic field, assumed to be perpendicular to the plane defined by the graphene. Now, the resultant Hilbert space for the dynamical states of any electron will be the tensor product of the dynamical spaces of each band under the application of $\mathbf{B}$, i.e., $\varepsilon^{\mathbf{B}} = \varepsilon^{\mathbf{B}}_{VB} \otimes \varepsilon^{\mathbf{B}}_{CB}$, where $\varepsilon^{\mathbf{B}}_{VB}$ and $\varepsilon^{\mathbf{B}}_{CB}$ are the corresponding dynamical spaces for the valence and conduction bands, respectively. (These dynamical states will be the same for both DP, -see above-.)

Therefore, due to the contribution of the two bands, under the application of the magnetic field there will be two degrees of freedom and, then, the quantized energy spectrum has to be described by that corresponding to an isotropic bidimensional harmonic oscillator. Hence, the quantized levels of electrons in graphene, in each K and K' points, will be given by the following expression

$$E_n = (i+j)\hbar\omega_0 = n\hbar\omega_0 = nE_0 \qquad (1)$$

with $i, j$, natural numbers related to each dynamical space, $n=0, 1, 2\ldots$, $\omega_0 = eB/m^*$ being the fundamental angular frequency, and $E_0 = \hbar\omega_0$.

The next step to give in the determination of the magntetotransport properties of graphene will be, in the context of the model, the calculation of the density of states. From equation (1), using the Possion summation formula, taking into account the spin-orbit interaction, can be obtained the expression for the density of states in each DP [4,13,14,26]

$$g^{DP}(E) = g_0 \left\{ 1 + 2\sum_{p=1}^{\infty} A_{S,p} A_{\Gamma,p} \cos\left[2\pi p\left(\frac{E}{\hbar\omega_0}\right)\right] \right\} \qquad (2)$$

where $p$ is the summation index corresponding to the $p$ harmonic, and $g_0$ is the two-dimensional density of states in absence of magnetic field.

$A_{S,p} = cos\left(\pi p \frac{g^*}{2} \frac{m^*}{m}\right)$ is the term associated with the spin and spin-orbit coupling, where $g^*$ is the generalized gyromagnetic factor, -that we assume here to be the same for both bands and with a value of 2, in correspondence with the experimental measurements of this magnitudes [22,27,28]-.

$A_{\Gamma,p} = exp\left\{-\frac{2\pi^2 p^2 \Gamma^2}{\hbar^2 \omega_0^2}\right\}$ is the term related to the width of the energy levels, associated directly with the factor $\Gamma$, due to the interaction of the electron with defects and ionized impurities in the system [29]. In all bellow, for sake of the simplicity, we assume constant gaussian width for every energy level.

The next step in the development of the model will be to include the two valley degeneracy, K and K', which corresponding spectra we suppose to be identical. Hence, the total density of states of the 2DES in MLG can be expressed by

$$g^{total}(E) = g^K(E) + g^{K'}(E) \qquad (3)$$

where the indexes refer to the density of states of each DP, given by equation (2).

### 3. The model for both magnetoconductivities

In this section we will initially consider the DP K. Our starting point will be the magnetotransport expressions for the 2DES in the semiconductor QW developed in the references by Hidalgo [4,11,14]. Then, for the diagonal magnetoconductivity at high magnetic fields we find

$$\sigma_{xx}^{K} = \frac{1}{\omega_0 \tau} \frac{en^{K}}{B} \frac{g^{K}(E_F)}{g_0} \qquad (4)$$

where $g(E_F)$ is the density of states at the Fermi level as obtained from equation (2), and $\tau$ represents the relaxation time of any electrons.

On the other hand, for the Hall magnetoconductivity at high magnetic fields we have

$$\sigma_{xy}^{K} = -\frac{en^{K}}{B} \qquad (5)$$

being $n^{K}$ the electron density at K, obtained from the density states, equation (2), and given by [4,13,14]

$$n^{K} = n_0 + \frac{2eB}{h} \sum_{p=1}^{\infty} \frac{1}{\pi p} A_{S,p} A_{\Gamma,p} A_{T,p} sen\left[2\pi p \left(\frac{E_F}{\hbar \omega_0}\right)\right] = n_0 + \delta n \qquad (6)$$

with $n_0$ the density of electrons at zero magnetic field (or zero gate voltage); and $\delta n$ the variation in the electron density as a consequence of the quantized density of states.

The effect of the temperature, which ultimately determines the occupation of every state, is incorporated in the term $A_{T,p} = z\, senh(z)$, where $z = 2\pi^2 pkT\, \hbar \omega_0$, $k$ being the Boltzmann constant and $T$ the corresponding temperature.

Equations (4) and (5) are the magnetoconductivities expressions of the model for K or K'. Now, because we are assuming the two DP to be equivalent we can add the corresponding magnetoconductivities, obtaining

$$\sigma_{xy}^{total} = \sigma_{xy}^{K} + \sigma_{xy}^{K'} = -\frac{en^{total}}{B} \qquad (7)$$

where $n^{total}$ is the total density of electrons, i.e. $n^{total} = n^{K} + n^{K'} = 2n^{K}$.

And for the diagonal magnetoconductivity,

$$\sigma_{xx}^{total} = \sigma_{xx}^{K} + \sigma_{xx}^{K'} = \frac{1}{\omega_0 \tau} \frac{en^{total}}{B} \frac{g^{total}(E_F)}{g_0} \qquad (8)$$

with the density of states at the Fermi level given by Equation (3). $\tau$ is the relaxation time of the electrons in each DP that we assume to be the same.

From equations (7) and (8) is immediate to calculate the symmetric magnetoresistivity tensor, $[\rho]=[\sigma]^{-1}$, whose terms are determined by the expressions

$$\rho_{xx} = \rho_{yy} = \frac{\sigma_{xx}}{\sigma_{xx}^2 + \sigma_{xy}^2} \qquad (9)$$

$$\rho_{xy} = -\rho_{yx} = -\frac{\sigma_{xy}}{\sigma_{xx}^2 + \sigma_{xy}^2} \qquad (10)$$

## 4. Simulations

This section is devoted to the analysis of the results of the model and its comparison with some experimental measurements. We show how the model performs as a function of the gate voltage and the magnetic field and, also, how is the evolution of the magnetotransport magnitudes with temperature it predicts.

From the evolution with temperature of the maxima of the SdH oscillations, Tiras et al. [25] have determined the effective mass for the electrons in the MLG, finding that has a value of the order of $m^* = 0.0124 m_0$ ($m_0$ being the free electron mass), in coincidence with other measurements in literature.

    a) GQHE as a function of the gate voltage:

For testing the model, firstly we simulate the experimental measurements in MLG as a function of the gate voltage.

Assuming a linear relation between the gate voltage and the Fermi level, we can rewrite this in equation (6) as

$$E_F = eV_g \qquad (11)$$

For an undoped graphene samples, because the Fermi level is initially in the Dirac point, we can assume $n_0=0$ in equation (6), and then applying a gate voltage $V_g>0$ the carrier population would be holes, and if $V_g<0$ electrons.

In Figure 1 we show the results obtained with the model for the Hall magnetoconductivity, (a), and the Hall and diagonal magnetoresistivities, (b); at the gate voltage interval between -80 and 80 V. To achieve the simulation we have assumed a temperature of 1.6 K; a gyromagnetic factor of 2; and a magnetic field $B=9$ T, (conditions of the measurements obtained by Zhang et al. [28]). The effective mass considered is $m^* = 0.0124 m_0$ [25]. Additionally, we have assumed for the simulation a constant Gaussian width with $\Gamma=0.06$ eV and a relaxation time $\tau=1$ ps. As it is clearly seen in the figures, the model reproduces accurately the experimental results, appearing the plateaux in the Hall magnetoconductivity at the values given by

$$\sigma_{xy} = \pm 2(2n+1)\frac{e^2}{h} \qquad (12)$$

in correspondence with the observed plateaux series ±2, ±6, ±10, ±14, ±18, ±22…

It is also seen a common characteristic of all the quantum Hall effects: the gate voltage intervals where the SdH oscillations are minima match with the plateaux ones.

For the sake of completeness, in Figure 2 it is shown the Hall magnetoconductivity as a function also of the gate voltage but for different magnetic fields, $B=9$ T, (a), 15 T, (b), and 25 T, (c), (the rest of the conditions assumed for achieving this simulation have been the same as in Figure 1).

    b) GQHE as a function of the magnetic field:

Although not many measurements are found in literature related to the magnetotransport properties of graphene as a function of the magnetic field, the model allows us to analyse them. In these cases the Fermi level will be fixed and given by [13,14,15]

$$E_F = \frac{\pi \hbar^2 n_0}{m^*} \qquad (13)$$

where $n_0$ is the electron density of the 2DES at zero magnetic field, (easily obtained from the Hall measurements at low magnetic fields).

Locating the magnetic field values of the experimental maxima of the SdH oscillations, we can easily see that they match with the magnetic fields given by the expression

$$B_n \approx n \frac{m^* E_F}{\hbar e} \qquad (14)$$

For what we have used equation (1). $n$ is a natural number, and $E_F$ is given by equation (13). In fact, it is easily verifiable that the index $n$ is related to the plateaux index $2(2n+1)$. (The same equation (14) can be used to predict the position of the maxima of the SdH oscillations in the experiments of the IQHE, once we know the electron density at zero magnetic field. Moreover, a similar expression between the quantized levels and the plateaux is found in the FQHE.[15])

In Figure 3 it is shown the simulations obtained with the model for the Hall magnetoconductivity, (a), and diagonal magnetoresistivities, (b), as a function of the magnetic field, and for a sample at a temperature of 1.6 K; with an electron density $n_0 = 6 \times 10^{15}$ m$^{-2}$; a gyromagnetic factor of 2; a constant Gaussian width for the energy levels with $\Gamma = 0.025$ eV; a relaxation time $\tau = 1$ ps; and an effective mass of $m^* = 0.0124 m_0$.

The plateaux appear at the right places in the entire magnetic field interval.

   c) Effect of the temperature on the GQHE:

Finally, we show how the model reproduces the effect of the temperature on the magnetotransport properties, incorporated though the term $A_{T,p}$ in equation (6). Unlike the IQHE and the FQHE in semiconductor QW, the GQHE has already been observed even at ambient temperatures [17]. Therefore, we have also tested the model under these conditions.

In Figure 4 we show the result of the simulation for the Hall magnetoconductivity as a function of the gate voltage for a temperature of T=300 K. We have fixed a magnetic field of $B$=29 T, just as used in the paper by Zhang et al. [28], being the rest of the conditions the same as in Figure 1. We see that the model is capable to explain the appearance of plateaux even at so high temperatures.

## 5. Summary and Discussion

In the present work we have described a single electron approach to analyse the GQHE. Making an easy extension of our previous model for the IQHE [13,14,15], it reproduces all the observed characteristics of the magnetotransport properties in the MLG; what additionally means that this phenomenon can be integrated in the same framework as the IQHE [13,14] and the FQHE [15]: the same first physical principles and similar hypothesis and assumptions.

The only distinguishing aspects are to impose the existence of the quantization of two bands, the VB and CB, -which lastly leads to equation (1)-, and the twofold valley degeneracy, -associated with the presence of the K and K' points in the first Brillouin zone-.

As it has been shown along the paper the model is capable, in a natural and simple way, of explaining the series of the Hall plateaux observed experimentally, and even their appearance at high temperatures. Therefore, we hope that it can be useful in understanding the conduction mechanisms and magnetotransport properties of MLG.

In forthcoming works we will analyse, in light of the model described here, the experimentally observed plateaux with index 0, 1 and 4; the behaviour of the diagonal magnetoconductivity at $V_g$ values close to zero; and, finally, it will be extended to include the magnetotransport properties observed in BLG.

**Acknowledgments:**

The author would like to thank R. Cangas for valuable discussions.

**Figure 1: Simulation with the model of the Hall magnetoconductivity, (a), and the Hall and diagonal magnetoresistivities, (b), of the monolayer graphene as a function of the gate voltage**: The gate voltage interval is -80 to 80 V. The conditions of the simulation are: a temperature of 1.6 K; a magnetic field of $B$=9 T [28], and a gyromagnetic factor of 2. The effective mass assumed is $m^* = 0.0124 m_0$ [25]. Besides, we have considered a constant Gaussian width with $\Gamma$=0.06 eV and a relaxation time $\tau$=1 ps. In figure (a) the integer numbers over the reference lines label the different plateaux.

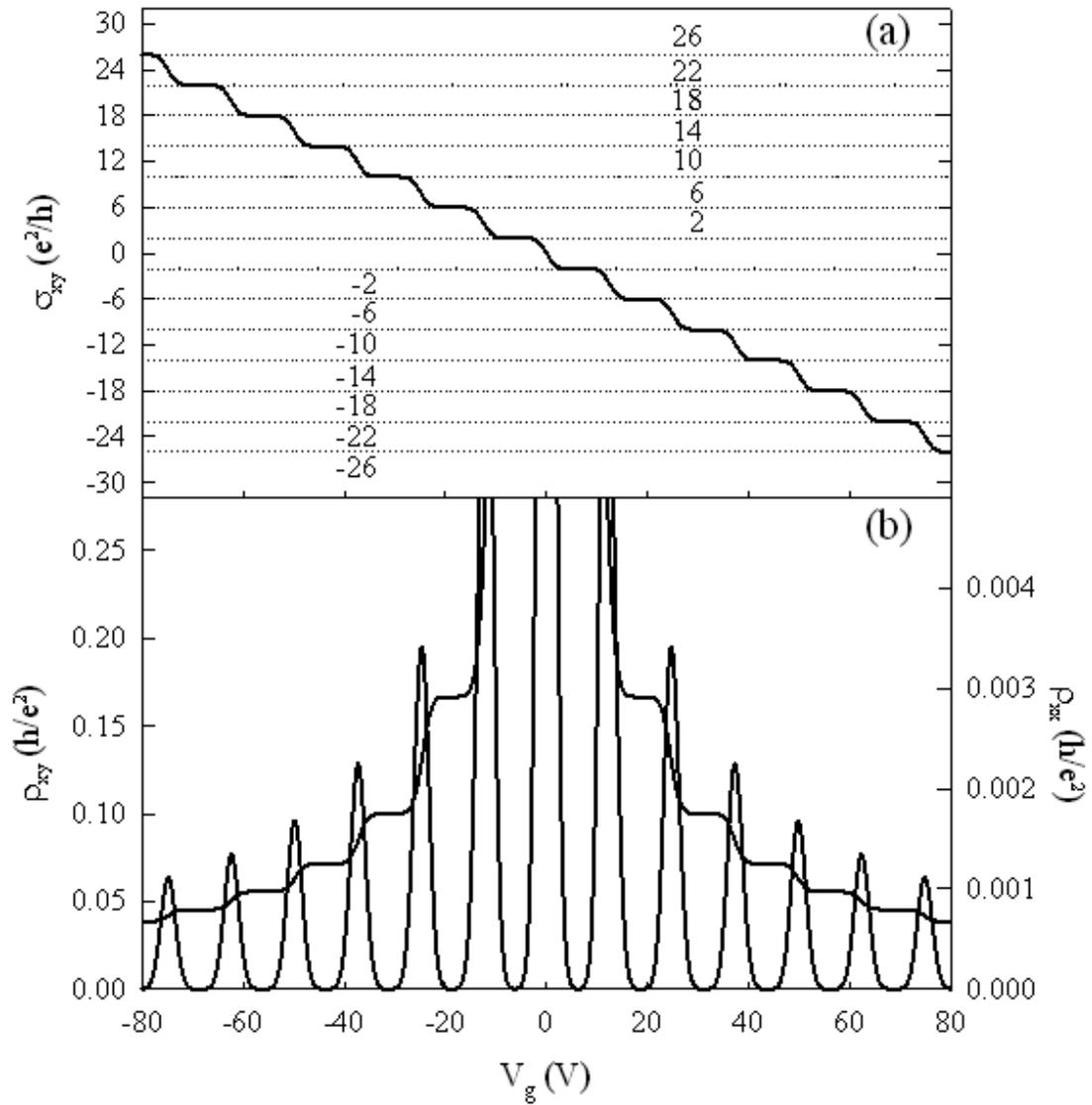

**Figure 2: Simulation with the model of the Hall magnetoconductivity for the monolayer graphene as a function of the gate voltage for different magnetic field values**: The magnetic fields fixed are $B$=9 T, (a), 15 T, (b), and 25 T, (c). The other physical conditions for the simulation are the same as in Figure 1. The integer numbers over the reference lines label the different plateaux.

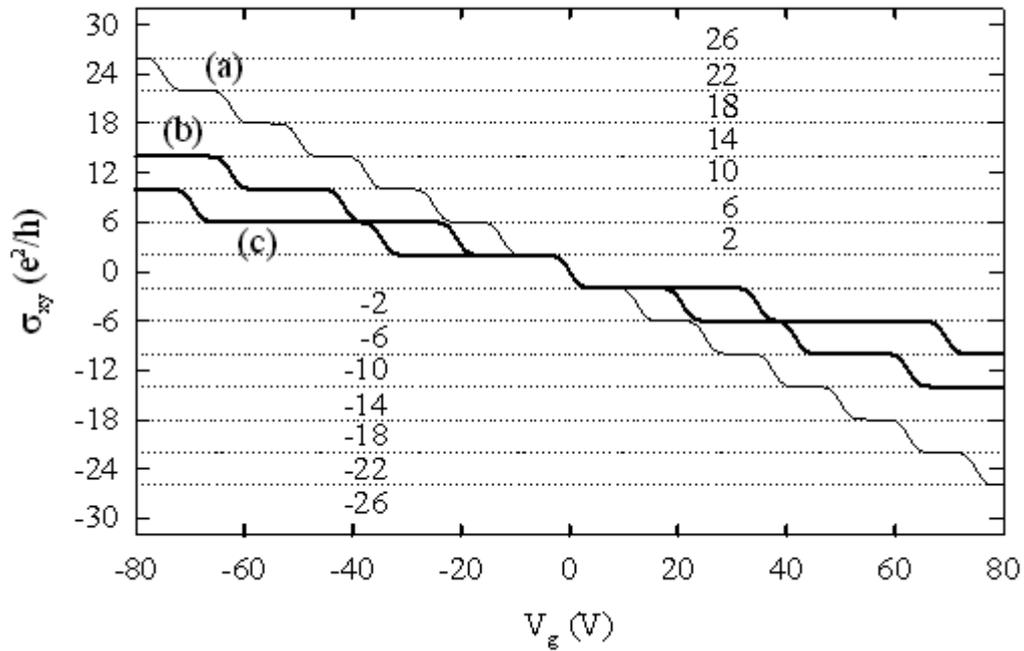

**Figure 3: Simulation with the model of the Hall magnetoconductivity, (a), and diagonal magnetoresistivities, (b), for the monolayer graphene as a function of the magnetic field**: To achieve the simulation we have considered a temperature of 1.6 K; an electron density $n_0=6\times10^{15}$ m$^{-2}$; a gyromagnetic factor of 2; and an effective mass of $m^* = 0.0124 m_0$ [25]. Besides, we have take into account a constant Gaussian width with $\Gamma=0.025$ eV and a relaxation time $\tau=1$ ps. In figure (a) the integer numbers over the reference lines label the different plateaux.

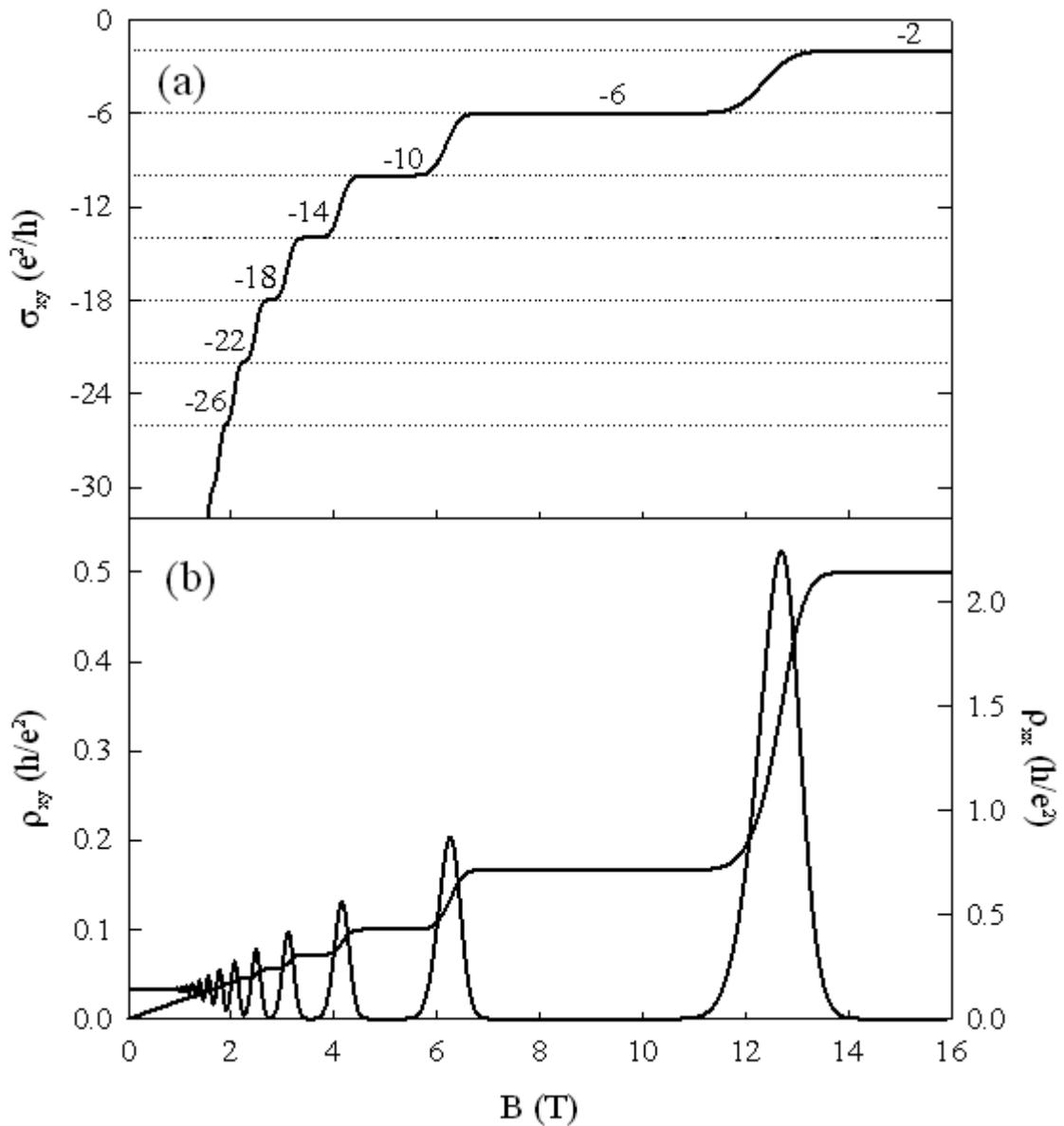

**Figure 4: Simulation with the model of the Hall magnetoconductivity for the monolayer graphene as a function of the gate voltage at a temperature of 300 K**: For the simulation we have considered $T$=300 K and a magnetic field $B$=29 T. The other fixed physical conditions are the same as in Figure 1. The integer numbers over the reference lines label the different plateaux.

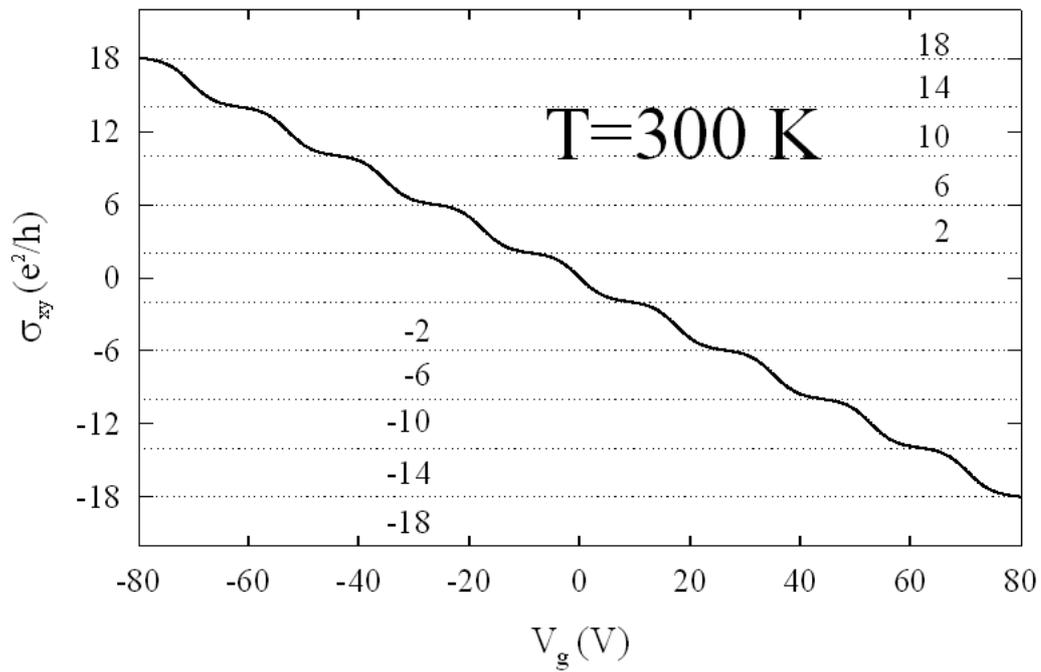